\documentstyle[prd,aps,eqsecnum,floats]{revtex}
\def\be{\begin{equation}}
\def\ee{\end{equation}}
\def\ba{\begin{eqnarray}}
\def\ea{\end{eqnarray}}

\def\negenspace{\kern-1.1em}

\begin{document}
%\draft
\input epsf

%@twocolumnfalse\endcsname
%\widetext

\title{Eigenvalues of the Stewart--Lyth equation for inflation
with a blue spectrum}

\author{Franz E. Schunck$^\$$\thanks{E-mail: fs@thp.uni-koeln.de}\\
$^\$$ Institut f\"ur Theoretische Physik,\\
Universit\"at zu K\"oln, 50923 K\"oln, Germany\\
and\\ Eckehard W. Mielke$^\diamond$\thanks{E-mail: ekke@xanum.uam.mx}\\ 
$^{\diamond}$ Departamento de F\'{\i}sica,\\
Universidad Aut\'onoma Metropolitana--Iztapalapa,\\
Apartado Postal 55-534, C.P. 09340, M\'exico, D.F., MEXICO\\
}

%\date{\today}

\maketitle

\begin{abstract}
By using the rather stringent {\em nonlinear}
second order slow--roll approximation, we reconsider the nonlinear
second order Abel equation of Stewart and Lyth. We determine a new
{\em blue} eigenvalue spectrum. Some of the discrete values of the
spectral index $n_s$ have consistent fits to the cumulative COBE
data as well as to recent ground-base CMB experiments.
\end{abstract}
\vspace{0.7cm}


%\narrowtext
%\vskip2pc]

%*********************************************************
\section{Introduction}

For a wide range of inflationary models \cite{lidlyt2}, 
a single scalar field, {\em the inflaton}, is presumed to be rolling in some 
underlying potential $U(\phi)$. This standard scenario is generically referred
to as {\em chaotic inflation}. A variance of it is the recently proposed 
{\em quintessence}. 

Our rather ambitious aim of reconstruction
is to employ observational data to deduce the complete functional form 
$U(\phi)$ of the inflaton self--interaction potential over the range 
corresponding to large scale structures, allowing a so--called graceful 
exit to the Friedmann cosmos \cite{Li95}.
The generation of density perturbations and gravitational waves has
been extensively investigated. The usual strategy is an 
expansion in the deviation from scale invariance, formally expressed as
the slow--roll expansion \cite{Cobl}.
     
In general, {\em exact} inflationary solutions, after an elegant 
coordinate transformation \cite{Sc94}, depend on 
the Hubble expansion rate $H$ as a new ``inverse time", and the 
regime of inflationary potentials allowing a graceful exit has already been 
classified \cite{MS95}.
In our more phenomenological approach, the inflationary dynamics is not 
prescribed by one's theoretical prejudice. On the contrary, in this 
solvable framework, the `graceful exit function' $g(H)$, which  
determines the {\em inflaton potential} $U(\phi)$ and 
the exact Friedmann type solution, is {\em reconstructed} in order to fit the
data. Even more, the transparent description of inflationary phase within the
$H$--formalism was the foundation for deriving a new mechanism of inflation,
called {\em assisted inflation} \cite{LMS98}.

Recent astronomical observations by COBE of the cosmic microwave background
(CMB) confirm that the Universe expands rather homogeneously on the large 
scale. From cumulative four years CMB observations, the spectral index is now 
measured by COBE as $n_{s}=1.2 \pm 0.3$, including the quadrupole ($l=2$) 
anisotropy \cite{bennett}.
This experiment has been complemented by ground-based CMB experiments. 
The preliminary data \cite{Scott} from the Cambridge Cosmic 
Anisotropy Telescope (CAT) e.g., are consistent with COBE, albeit a 
{\em slightly higher} spectral index of $n_s =1.3\pm 0.4$. This allows 
\cite{LJD98} to constrain cosmological parameters, such as matter density
$\Omega_{\rm m}= 0.32 \pm 0.08$ and the Hubble constant to
$H_0=47 \pm 6$ km/(s Mpc). A more recent analysis \cite{DK99} of the CMB
anisotropy strongly constrain the
spatial curvature of the Universe to near zero, i.e.~$k\cong 0$.  
All data imply that a total density $\Omega =1$ (or $k=0$) is
$2\times 10^7$ more probable then $\Omega \cong 0.4$. Nowaday's best-fit to
{\em all} CMB data is a Hubble constant of $H_0=65$ km/(s Mpc) and requires
{\em dark energy} with $\Omega_\Lambda=0.69$ and a
spectral index $n_s=1.12$.

Stewart and Lyth \cite{stelyt} proceeded from the exact power--law 
inflation in order to analytically compute the second order slow--roll
correction to the standard formula \cite{ref}.
These analytical results are at present the most accurate equations available. 
This remarkable accuracy has recently been confirmed \cite{GL96} by numerical 
perturbations of the exact analytical result for the power--law inflation.

We apply the $H$--formalism to this accurate second order perturbation 
formalism, and transform the nonlinear equation of Stewart and Lyth into an
{\em Abel equation}. In extension of our earlier work \cite{mom}, we 
determine here a {\em discrete}
{\em eigenvalue spectrum} of this nonlinear equation. In particular, our 
new {\em blue} discrete eigenvalue spectrum has the
scale invariant Harrison--Zel`dovich spectrum with index $n_s=1$ as a 
{\em limiting point}. Thus some specific eigenvalues $1<n_s< 1.5$ of the
family of solutions can be related
rather accurately to existing as well as future observations.

%******************************************************
\section{General metric of a spatially flat inflationary universe}

In the following we reconsider the Lagrangian density of a 
rather general class of inflationary models in the Einstein frame:
\be
{\cal L}_{\rm E} = \frac{1}{2 \kappa } \sqrt{\mid g \mid}
 \left( R+ \kappa \left[ g^{\mu \nu }(\partial_\mu \phi)(\partial_\nu 
 \phi) - 2 U(\phi) \right] \right)  \; .
\label{lad}
\ee
Here $R$ is the curvature scalar, $\phi $ the scalar field, $U(\phi)$
{\em the self--interaction potential} and $\kappa=8\pi G/c^{4}$ the
gravitational coupling constant. We use natural units with $c=\hbar =1$ and 
our signature for the metric is $(+1, -1, -1, -1)$ as it is common in particle
physics.

For the {\em flat} ($k=0$ or $\Omega=1$) Robertson-Walker metric favored by
the inflationary paradigm and by recent observations, the evolution of the
generic inflationary model (\ref{lad}) is
determined by the autonomous first order equations
\be
\dot H = g(\phi, H)\, , \qquad 
  \dot \phi = \pm \sqrt {2/\kappa}
  \sqrt{-g(\phi, H)}\, , \label{aut}
\ee
where $g(\phi, H) :=\kappa U(\phi ) - 3H^2$. De Sitter inflation 
with $\dot H=0$ is a {\em singular}, but well--known subcase, cf.~\cite{Ku95}.
For $g(\phi, H)\neq 0$, however, we can simply {\em reparametrize} the
inflationary potential $U(\phi )
\rightarrow \widetilde U = \widetilde U (H) := U(\phi(t(H)))$ in terms of
the Hubble expansion rate $H:=\dot a(t)/a(t)$ as the {\em new}
``inverse time" {\em coordinate}. For this non-singular coordinate 
transformation, the term $g(\phi, H)$ reduces to 
the graceful exit function
$g(H)=\kappa \widetilde U (H) -3 H^2 <0$ and the following general
metric and scalar field solution of (\ref{aut}) emerges \cite{Sc94,MS95}:
\ba
ds^2 & = &
\frac {dH^2}{g^2(H)}
 - a_0{}^2 \exp \left ( 2
   \int \frac {H dH}{g(H)} \right ) \left [ d r^2 + r^2 \left (
    d \theta^2 + \sin^2 \theta d \varphi^2 \right ) \right ] \; , \\
\phi & = & \phi (H) = \mp \sqrt {\frac{2}{\kappa }}
 \int \frac {dH}{\sqrt{-g(H)}} \; .
\label{phiH}
\ea
Once we know $g(H)$ from some experimental input, we can (formally) 
invert (\ref{phiH}) in order to recover from the chain of substitutions
\be
[g(H) +3H^2]/\kappa  =\widetilde U (H)=
\widetilde U(H\circ\phi^{-1}(\phi)) =
 U(\phi) 
\label{ipot}
\ee  
the inflaton potential in a simple and elegant manner.

%***********************************************************
\section{Second-order slow--roll approximation in canonical Abel form}

For a reconstruction of the potential $U(\phi)$ under the ``umbrella" of 
{\em chaotic inflation}, we adopt here the equations of Stewart and Lyth 
\cite{stelyt} for the {\em second order} slow--roll approximation. 
In terms of the energy density $\epsilon = -g(H)/H^{2}$, this {\em nonlinear} 
second order differential equation involving the {\em scalar} spectral
index $n_{s}$ reads \cite{mom,Mi98} 
\be
2 C \epsilon\stackrel{\cdot \cdot}{\epsilon} -
(2C+3)\epsilon\stackrel{\cdot}{\epsilon}
- \stackrel{\cdot}{\epsilon}
+ \epsilon^2  + \epsilon + \Delta  = 0  \; ,
\label{diff2}
\ee
where $\cdot \hat= d/d\ln H^{2}$, $C:=-2+\ln 2 +\gamma \simeq -0.73$, and 
\be
\Delta:= (n_{s}-1)/2
\ee
denotes the deviation from the 
scale invariant Harrison--Zel`dovich spectrum. 
The corresponding second order equation for the spectral index $n_{g}$ of 
{\em tensor} perturbations turns out to be
\be
n_{g}=-2\epsilon \left[1+ \epsilon - 2(1+C)
\stackrel{\cdot}{\epsilon}\right]\simeq n_s-1 \, , \label{tensor}
\ee
where the approximation is valid in first order.
Note that for the constant solution $\epsilon =A_0$, where
\be
A_0^2 + A_0 + \Delta = 0 \; , \quad \Rightarrow \quad 
A_0= {1\over 2} \left(-1\mp \sqrt{3-2n_{s}}\right)\, ,  \label{zero}
\ee
we recover from the second order equation (\ref{diff2}) exactly the {\em first
order consistency relation} $n_g=n_s-1$.
Moreover, we can infer already from (\ref{zero})
that a {\em real} eigenvalue spectrum is constrained by $n_s \leq 1.5$.

By introducing the flow of the energy density $\epsilon$ via
$p(\epsilon):= \stackrel{\cdot}{\epsilon}$ with 
$\stackrel{\cdot \cdot}{\epsilon}= p dp/d\epsilon$, the second order equation 
(\ref{diff2}) can be brought to the following form:
\be
\epsilon p {dp\over d\epsilon} -
(\beta+\alpha \epsilon) p  
+ \beta (\epsilon^2  + \epsilon +  \Delta) = 0  \; ,
\label{diff4}
\ee
where $\alpha=1+3/(2C)$ and $\beta=1/(2C)$.
For invertible energy flow, it can be transformed via $u :=1/p$ 
into the canonical form of the {\em Abel equation} of the first kind,
cf.~\cite{DLP98},
\be
2 C \epsilon {du\over d\epsilon} +u^2[1+
(2C+3)\epsilon]  -u^3\left[
 \epsilon^2  + \epsilon +  \Delta\right] = 0  \; .
\label{Abel}
\ee

Garc\'{\i}a et al.~\cite{Gr97} have demonstrated that, within the class of
{\em finite} polynomials  $p=\sum_{i=0}^N b_i \epsilon^i$,
the unique solution of (\ref{diff4}) is given by a polynomial 
$p=\Delta + b_1 \epsilon$ of {\em first} degree,
where $b_1= (\alpha \pm \sqrt{\alpha^2-4\beta})/2$ necessarily satisfies the
additional constraint
$b_1(1-b_1) =\Delta$ due to the {\em nonlinearity} of the Abel equation. 
Mathematically, this transformation improves the order of
the polynomial by one in comparison with (3.1), where the solution is
just a polynomial of zero degree. This first order polynomial corresponds 
to $\epsilon = A_0 +\epsilon_0 y^{1+A_0}$, where $A_0$ is again given by 
(\ref{zero}). Since $\Delta$ is determined due to the 
additional constraint mentioned above, 
cf.~\cite{bmmov}, the spectral index $n_{s}$ is now {\em completely fixed}  
by the Euler constant $\gamma$ as
\ba
n_{s}&=& 1 - \frac{2C + 9 \mp 3\sqrt{4C^{2} + 4C +9}}{4C^{2}}\nonumber \\
&=&1.49575 \quad (-6.57797) \, . \label{ns}
\ea
In order to circumvent this `no-go' theorem for polynomial solutions, 
we shall introduce
a different transformation which leads us to new solutions in form of an 
{\em infinite} series.

%*****************************************
\subsection{New class of solutions with a blue spectrum}

If we assume that $\epsilon=\epsilon(y)$ with $y:= H^2$, Eq.~(\ref{diff2})
can be rearranged to
\be
2 C \epsilon y^2 {\epsilon}''  -
(3\epsilon +1) y {\epsilon}' 
+ \epsilon^2  + \epsilon + \Delta  = 0  \; ,
\label{diff2a}
\ee
where ${}'\hat=d/dy$. We try the ansatz
\be
\epsilon=\sum_{i=0}^\infty A_i y^i  \label{ser} \; ,
\ee 
and find that the odd powers have to vanish. This can be traced back to 
the fact 
that (\ref{diff2a}) is {\em invariant} under reflections $y\rightarrow -y$ 
only for {\em even} functions 
$\epsilon(-y)= + \epsilon(y)$. For  even powers,
the following Taylor expansion can be recognized:
\be
\epsilon=\sum_{j=0}^\infty A_{2mj} y^{2mj}  \label{ser2} \; ,
\ee
where $m$ is a positive integer, which labels the first non-constant term 
in our expansion and thus distinguishes the {\em different subclasses} of 
solutions. Hence, the series for $m=1$ starts as
$\epsilon=A_0+A_2 y^2+A_4y^4+\ldots$, for the case $m=2$ as
$\epsilon=A_0+A_4 y^4+A_8y^8+\ldots$, and so on. In all cases, the coefficient
$A_0$ of zeroth order ($y^0$) is again determined algebraically by a $\Delta$
via Eq.~(\ref{zero}); or, as we shall show below, $A_0$ is determined by the
next higher order, and the zeroth order establishes $n_s$.

In order $2m$, we find the following {\em discriminating relation}
\be
\{[2m(2m-1)C- 3m +1]A_0-m+1/2\} A_{2m} = 0 \; ,
\ee
due to the {\em nonlinearity} of the Abel type equation. This recursion 
relation implies $A_{2m}=0$ (and $\epsilon=A_0=constant$) or the
{\em new additional} constraint
\be
A_0=\frac{2m-1}{2[2m(2m-1)C- 3m +1]} <0 \; .
\ee
Together with (\ref{zero}) this implies that 
the spectral index of our new class of solutions can only adopt 
the following {\em discrete eigenvalues}
\be
n_s=1+\frac{(2m-1)[4m -1 -4m(2m-1)C]}{2[2m(2m-1)C- 3m +1]^2} \,.
\label{eigenv}
\ee
For solutions with large $m$, we find asymptotically
\be
A_0 \simeq \frac{1}{4mC} \; , \quad
n_s \simeq 1 - \frac{1}{2mC} \; ,
\ee
such that the 
scale invariant Harrison--Zel`dovich solution with index $n_s=1$ turns 
out to be {\em limiting point}. Let us stress that 
our new discrete spectrum approaches it from the 
{\em blue} side, which previously was considered rather difficult to 
achieve.
In order to correlate this with observational restrictions, let us 
display the highest eigenvalues for $n_s$. For $m=1$, we have
$(4CA_0-4A_0-1) A_2 = 0$ and, because of $A_2\neq0$,
\be
A_0 = \frac{1}{4(C-1)} = -0.1445 \; .
\ee
Eq.~(\ref{zero}) discovers the value of the scalar spectral index to be
\be
n_s = 1.247 \; . \label{m1}
\ee
The integration constant $A_2$ remains arbitrary. The fourth order establishes
the constant $A_4$ as an algebraic function of $A_0,A_2$ and similar for
higher orders.

For $m=2$, the fourth order constraint is
\be
A_0 = \frac{3}{24C-10} = -0.109 \; ,
\ee
while $A_4$ now plays the role of a free constant. Then, zeroth order
vanishes if
\be
n_s = 1.194 \; ,\label{m2}
\ee
which is very close to the experimental value $n_s =1.2 \pm 0.3$ from  
COBE. Furthermore, it is revealed that $A_8=f_1(A_4,A_0)$,
$A_{12}=f_2(A_8,A_4,A_0)$, and so on (where $f_1,f_2$ are some algebraic
relations).
The higher order coefficients of the series can be derived recursively via
{\sc Mathematica}, similarly as in the case of the Bartnik--McKinnon solution, 
cf.~\cite{Sch93}.

%************************************************************************

\section{Reconstruction for the full eigenvalue spectrum} 

In cosmological applications, the energy density $\epsilon$ in our second
order reconstruction has only a limited range.
For the decelaration parameter
$q(t):=- \stackrel{\cdot \cdot}{a}a/ \stackrel{\cdot}{a}{}^2$ $= \epsilon -1$ 
to become negative, as necessary for inflation, our solutions are 
constrained by $0< \epsilon <1$. In the case of the constant solution 
$\epsilon= A_0$, merely the spectral index is restricted to the 
{\em continuous} 
range $-3 <n_s< 1$. Since this corresponds to $g(h) =-A_0 H^2$, a simple 
inversion of the integral (\ref{phiH}) and subsequent insertion into 
(\ref{ipot}) leads to the potential 
\be
  U(\phi ) = \frac {3-A_0 }{\kappa } \tilde C^2
 \exp \left(\pm \sqrt{2 \kappa A_0 }\; \phi \right)
\label{cobep}\, ,
\ee
of {\em power--law inflation}, where $\tilde C$ is an integration constant.

Similarly, for the extreme blue spectrum (\ref{ns}) the graceful exit 
function reads 
$g(H) =-H^2\left[A_0 +\epsilon_0 H^{2(1+A_0)}\right]$
yielding via (\ref{phiH}) and (\ref{ipot}) the 
reconstructed potential
\be
U(\phi) = {1 \over \kappa}  \left[-{A_0 \over \epsilon_0} \left(
1 + \tan^2 (P\phi) \right) \right]^{1/(1+A_0 )}
\left( 3 + A_0  \tan^2 (P\phi) \right) \, ,
\label{cobpot}  
\ee
where $P = \pm \sqrt{-\kappa A_0 /2\,}\,(1+A_0 )$ is a constant.
According to its shape, cf.~\cite{bmmov},
it interpolates between the power--law inflation in the limit 
$\epsilon_0\rightarrow 0$ and intermediate inflation. 
Since $n_{s}\simeq 1.5$, the parameters are approximately
$A_0 \simeq {-1/2}$ and $P\simeq \sqrt{\kappa}/4$. 
Its form is similar to the exact solution of Easther \cite{ea96} except 
that the latter yields the too large value $n_{s}=3$ for the spectral index.

\begin{figure}
\centering \leavevmode \epsfxsize=8cm \epsfbox{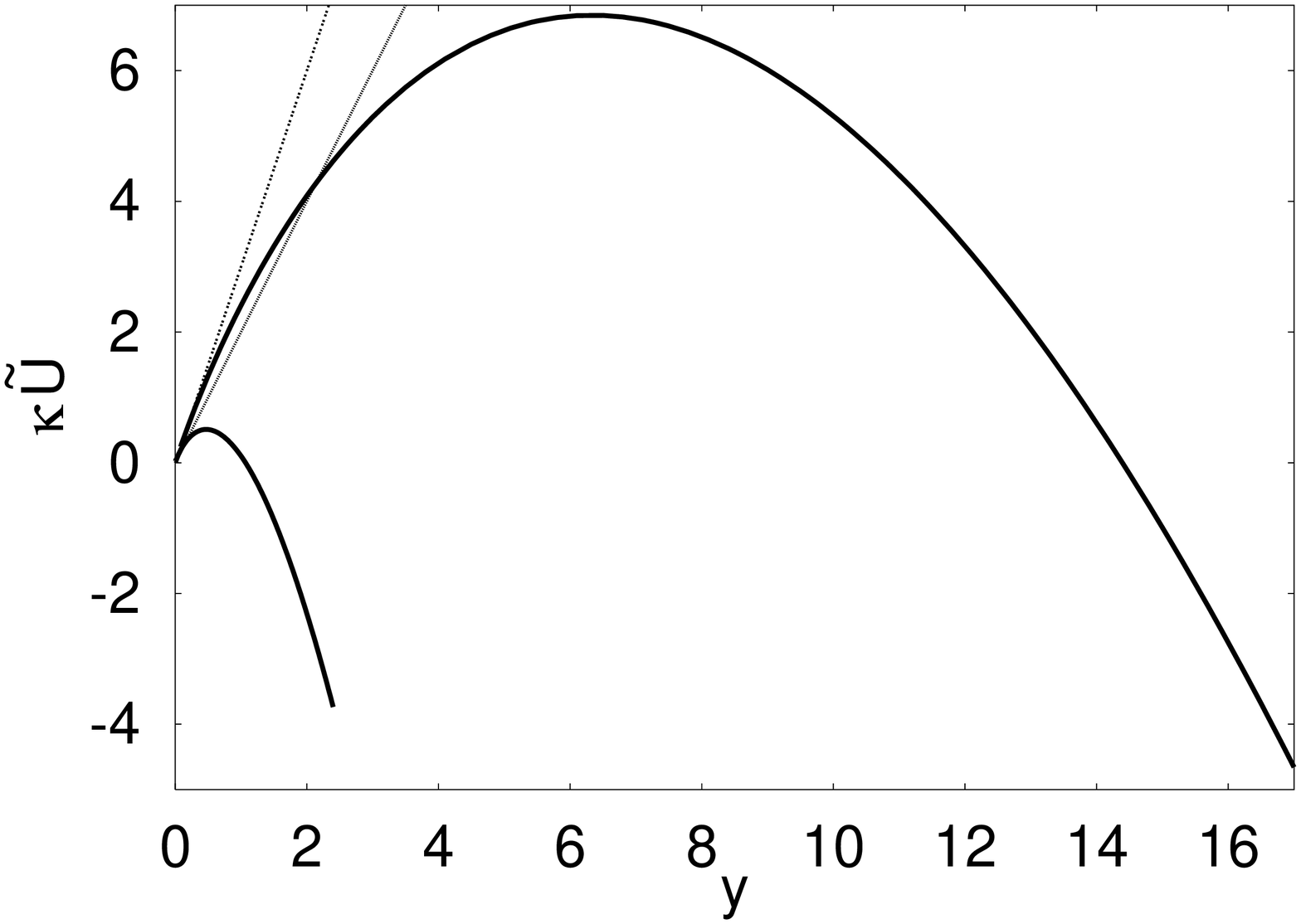} \hfill \\
\vskip.5cm
\caption[]{Inflationary potential $\kappa\widetilde  U(y=H^2)$ for $m=1$
  (potential function 
until $y=2.4$) and $m=2$ (until $y=17$). The two lines correspond to
$ \kappa\widetilde U=2y$ and $\kappa\widetilde  U=3y$, the boundaries of the
inflationary phase. Further explanation in the Appendix.}
\label{potH1}
\end{figure}

\begin{figure}
\centering \leavevmode \epsfxsize=8cm \epsfbox{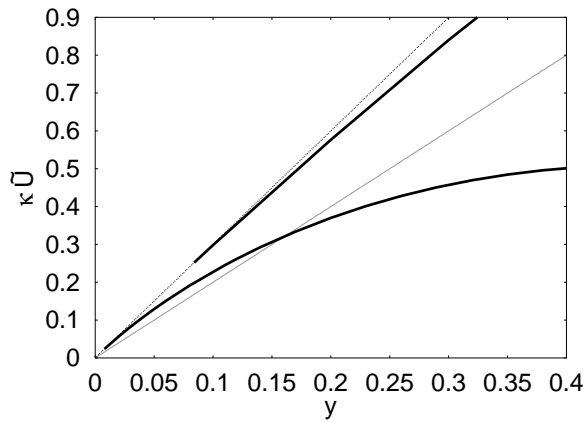} \hfill \\
\vskip.5cm
\caption[]{Enlarging the surrounding of the origin of Fig.~\ref{potH1}.
It is recognizable that both potentials start at the line 
$\kappa\widetilde  U=3y$, below the
boundary where real scalar fields are present. For the ($m=1$)-solution,
the initial value $\bar y$ is approximately $0.008026$ (for the values given
in the Appendix), and for $m=2$, $\bar y \approx 0.083427$.}
\label{potH2}
\end{figure}

For the new {\em blue} spectrum (\ref{eigenv}) we have  
reconstructed the potential $U(\phi)$ numerically in Fig.~\ref{pot1} and
\ref{pot2} for the first two indices (\ref{m1},\ref{m2}), see the Appendix
for details. The form of both new potentials resemble the ones found in
\cite{ea96,mom}. The corresponding forms of $\kappa\widetilde U(H^2)$
are shown in
Fig.~\ref{potH1} and \ref{potH2}. Inflation occurs if the scalar field is
between the two lines $2H^2$ and $3H^2$, cf.~\cite{Sc94}.

\begin{figure}
\centering \leavevmode \epsfxsize=8cm \epsfbox{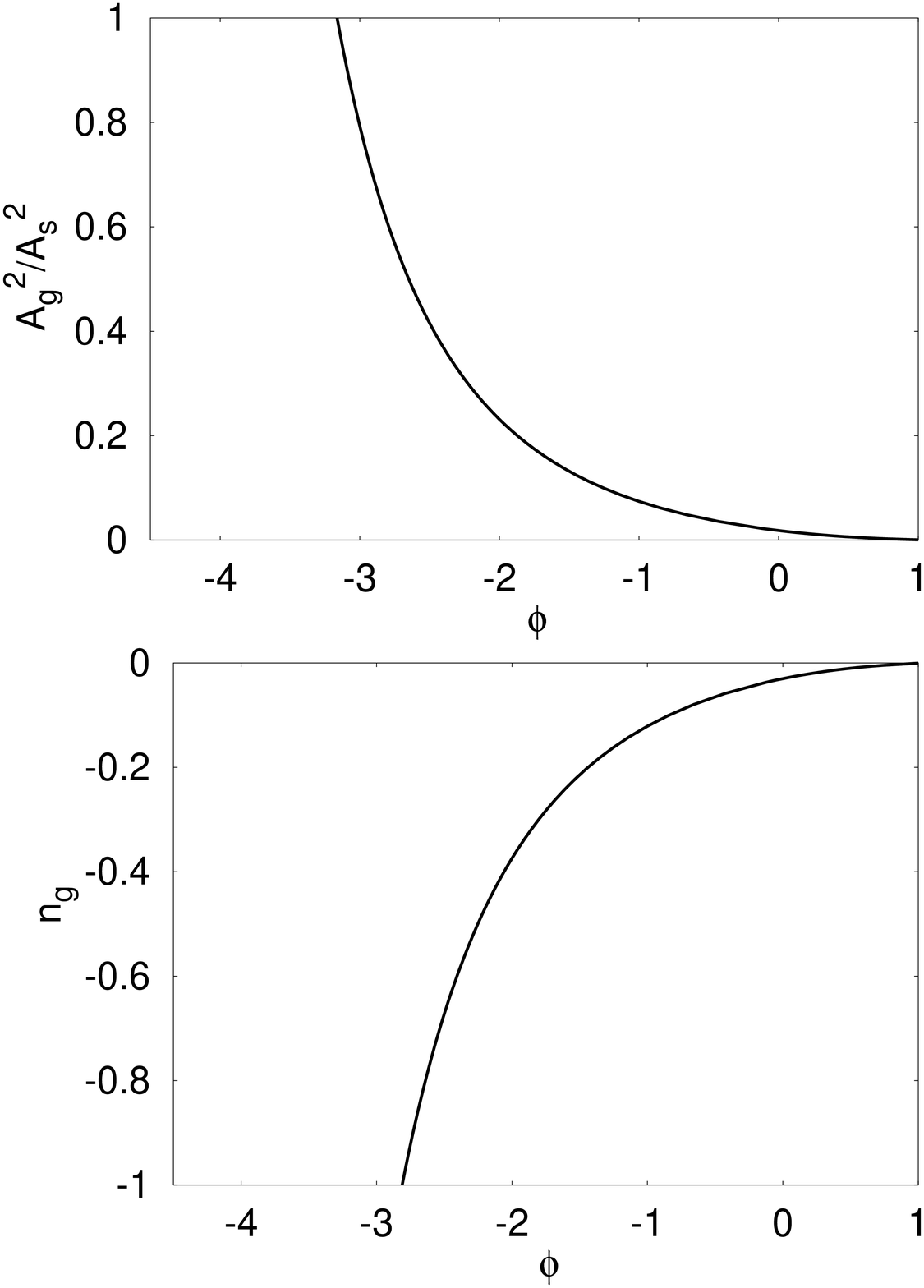} \hfill \\
\vskip.5cm
\caption[]{The ratio of the tensor and the scalar spectrum $A_g^2/A_s^2$
and the tensor spectral index $n_g$ for the ($m=1$)-solution within
inflationary region.}
\label{ng1}
\end{figure}

\begin{figure}
\centering \leavevmode \epsfxsize=8cm \epsfbox{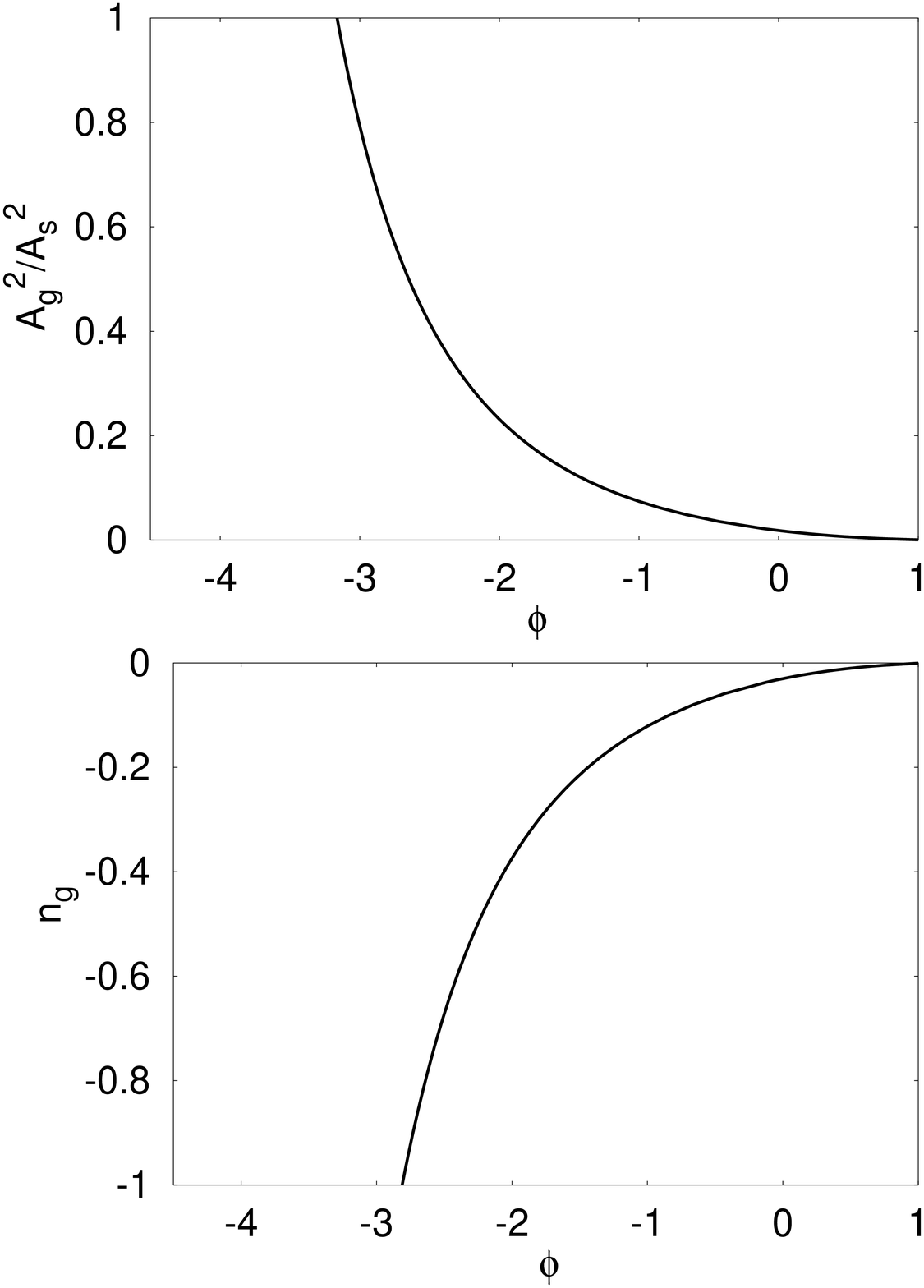} \hfill \\
\vskip.5cm
\caption[]{The ratio of the tensor and the scalar spectrum $A_g^2/A_s^2$
and the tensor spectral index $n_g$ for the ($m=2$)-solution within
inflationary region.}
\label{ng2}
\end{figure}

The range of $\phi$ for which these potentials are valid depends also on  
the (so-far missing) information on the spectral index
$n_g$ of tensor perturbations, cf.~Eq.~(\ref{tensor}). For our solutions
for $m=1$ and $m=2$ we show the $\phi$ dependence of $n_g$ in
Fig.~\ref{ng1} and \ref{ng2}. Since all our new solutions are 
`deformations' of the constant solution $\epsilon =A_0$, we expect the first
order consistency relation of power-law inflation 
to hold to some approximation, i.e.~$n_g\simeq n_s-1\in (-0.4, 0.5]$
\cite{k96}. In second order, we recognize from Fig.~\ref{ng1} and
\ref{ng2} that for our solutions $n_g$ is always negative.

Of particular interest \cite{k96} is also the squared ratio of the 
two further observables,
the amplitude $A_g$ of gravitational waves, versus its counter part $A_s$, 
the amplitude of primordial scalar density perturbations:
\be
\left(A_g^2/A_s^2 \right ) = \epsilon \left[1 - 2C
\stackrel{\cdot}{\epsilon}\right] \simeq {1\over 2} 
\left(\sqrt{3-2n_{s}}-1\right) \; ,
\ee
where the approximation is valid in first order.
We display the numerical results in Fig.~\ref{ng1} and \ref{ng2} as well.
Form first order approximations, we expect the range of this ratio within
$\sim [0,0.2)$. For several well-known inflationary models
(power-law and polynomial chaotic, e.g.) the predicted ratio can be found in
Table 2 of \cite{k96}. Recently, assisted
inflation was able to produce a scalar spectral index closer to
scale-invariance (but for $n_s<1$) because of its large number of
non-inflationary real scalar fields present in this theory \cite{LMS98}.

In this paper, we followed the basic idea of reconstructing the inflaton
potential in second
order, given in \cite{Li95}. But, instead of producing the
potential just for a few values $\phi$, we could construct the complete
functional form of the potential for a series of constant spectral indices.
As it is well-known, a full reconstruction can only be achieved if
informations on the gravitational parts are included; scalar perturbations
determine the potential up to an unknown constant. Since the
differential equations are nonlinear, different constants yield 
different potentials but for the identical scalar spectrum.
Any piece of knowledge
concerning the tensor perturbations breaks this degeneracy; so
we use our solutions to calculate both the tensor index $n_g$ and
the ratio of the 
amplitudes $A_g^2/A_s^2$ necessary in second order to define the
inflaton potential uniquely. We recognize that both parameters are not
constant in contrast to the scalar spectral index. We have checked numerically
several consistency equations in second order and found good
confirmation within the inflationary region. Furthermore, we analyzed
the behavior of the slow-roll parameters $\epsilon$ and $\eta=-dg/dH^2$,
cf.~the definition in \cite{Li95,MS95}. At the beginning of inflation,
we derive that $\eta>2\epsilon$ which is valid for the ($m=1$)-solution
up to $\epsilon \simeq 0.32$ and for the ($m=2$)-solution up to
$\epsilon \simeq 0.28$. This feature resembles that of {\em hybrid inflation}
\cite{hybrid},
cf.~Table 1 of \cite{Li95}, where the influence of gravitational waves is
negligible in accordance  with the small values of both $n_g$ and
$A_g^2/A_s^2$ as can be recognized within Fig.~\ref{ng1} and \ref{ng2}
close to $\phi \simeq 1$.

Summarizing, the nonlinear Abel equation (\ref{diff2}) permits exact 
solutions for the discrete or continuous eigenvalues of Table \ref{table}.
All positive discrete values are marginally consistent with the
$n_s =1.2 \pm 0.3$ from COBE as well as recent ground-based CMB experiments,
our highest value for the blue spectrum is just below
the $n_s\leq 1.5$ constraint in order to suppress distortions of the
microwave spectrum and the formation of primordial black holes during the
phase of reheating, cf.~\cite{Li95}.

\section{String or $M$ theory axion as quintessence}

Recently, {\em quintessence} in the form of slow-rolling scalar field $Q$
has been proposed as an alternative form of dark energy density with negative
pressure \cite{quint}, and Refs.~therein.
The true minimum of the quintessence potential is presumed to vanish,
i.e.~$(  U_{\mathrm Q})_{\rm min}=0$, 
however the present value of $Q$ is displaced from the true minimum,
providing 
\be 
  U_{\mathrm Q}\sim (3\times 10^{-3} \, {\rm eV})^4,
\label{quintpotential}
\ee 
and a negative pressure with the equation of state
\be 
\frac{p_{\mathrm Q}}{\rho_{\mathrm Q}}=
\frac{\dot{Q}^2-2  U_{\mathrm Q}}{\dot{Q}^2+2  U_{\mathrm Q}} \leq -0.6  \, ,
\ee 
while obeying the usual equations of motion.

Our potential (\ref{cobpot}) for the blue spectrum (\ref{ns}), mimics
to some extend the required properties of quintessence, cf.~\cite{BCN99}.

Given the assumption of $(U_{\mathrm Q})_{\rm min}=0$,
one can in principle compute the quintessence potential $  U_{\mathrm Q}$ 
once the particle physics model for $Q$ is given.
However, the required size of $U_{\mathrm Q}$ is extremely small compared
to any of the mass scales of particle physics,
and thus an utmost question is how such a small $  U_{\mathrm Q}$
can arise from realistic particle physics models.
In fact, the most natural candidates for a light scalar field whose typical
range of variation is of order $M_{\rm Planck}/\sqrt{8\pi}$ form string or
$M$-theory moduli multiplets describing the (approximately) degenerate string
or $M$-theory vacua \cite{LW99}.
It is thus quite tempting to look at the possibility that $Q$
corresponds to a certain combination of
the string or $M$-theory moduli superfields.

Current ground-based CMB experiments are already 
discriminating between different inflationary models 
and are preliminarily
constraining the cosmological parameters. In future, the
satellite--based microwave background anisotropy measurements from
{\em PLANCK Surveyor} will constrain
\cite{Li98} them even more, likely to less then 1 \%.

\acknowledgments
We would like to thank Ed Copeland, Andrew Liddle, Alfredo Mac\'{\i}as,
and Anupam Mazumdar for useful comments and hints to the literature.
This work was partially supported by CONACyT, grant No. 28339--E, the
grant P/FOMES 98-35-15, and the
joint German--Mexican project DLR--Conacyt
E130--1148 and MXI 010/98 OTH. One of us (E.W.M.) would like to
thank Noelia M\'endez C\'ordova for encouragement.

\appendix

\section*{Derivation of the new potentials}

Within the $H$-formalism, {\em real} scalar fields can only be determined if
$\kappa \widetilde U(H) < 3H^2$ \cite{Sc94,MS95}. Furthermore, inflation just
occurs within the area $2H^2 < \kappa \widetilde U(H) < 3H^2$.
We use this knowledge to calculate numerically the complete potential in
real $\phi $ space.
For instance, for $m=1$ we define a new function $f(y)$ by
\be
\epsilon =: A_0 + f(y) y^2
\ee
so that its corresponding differential equation
has the initial values $f(0)=A_2$ and $f'(0)=0$. The regime of real solutions
can be derived from the differential equation for the scalar field
\be
\phi' = \mp \sqrt {\frac{2}{\kappa }}
\frac {1}{\sqrt{ 4\epsilon y^2 }} \; .
\ee
Because $A_0<0$ we shall find an area around $y=0$ where $\phi$ is imaginary,
hence, $\kappa\widetilde U(H) > 3H^2$. The point defined by
$\bar y=\sqrt{-A_0/f(\bar y)}$,
corresponding to $\kappa\widetilde U(H) = 3H^2$, establishes then the initial
value $f(\bar y)$ (if $\bar y$ is chosen) in real $\phi $ space.
$f(\bar y)$ is still connected with the initial value $A_2$ in complex
$\phi $ space. The potential can be obtained immediately from the
differential equation
\be
\kappa\widetilde U' = (3-A_0) - y^3 f' - 3 y^2 f  \; .
\ee

\begin{figure}
\centering \leavevmode \epsfxsize=8cm \epsfbox{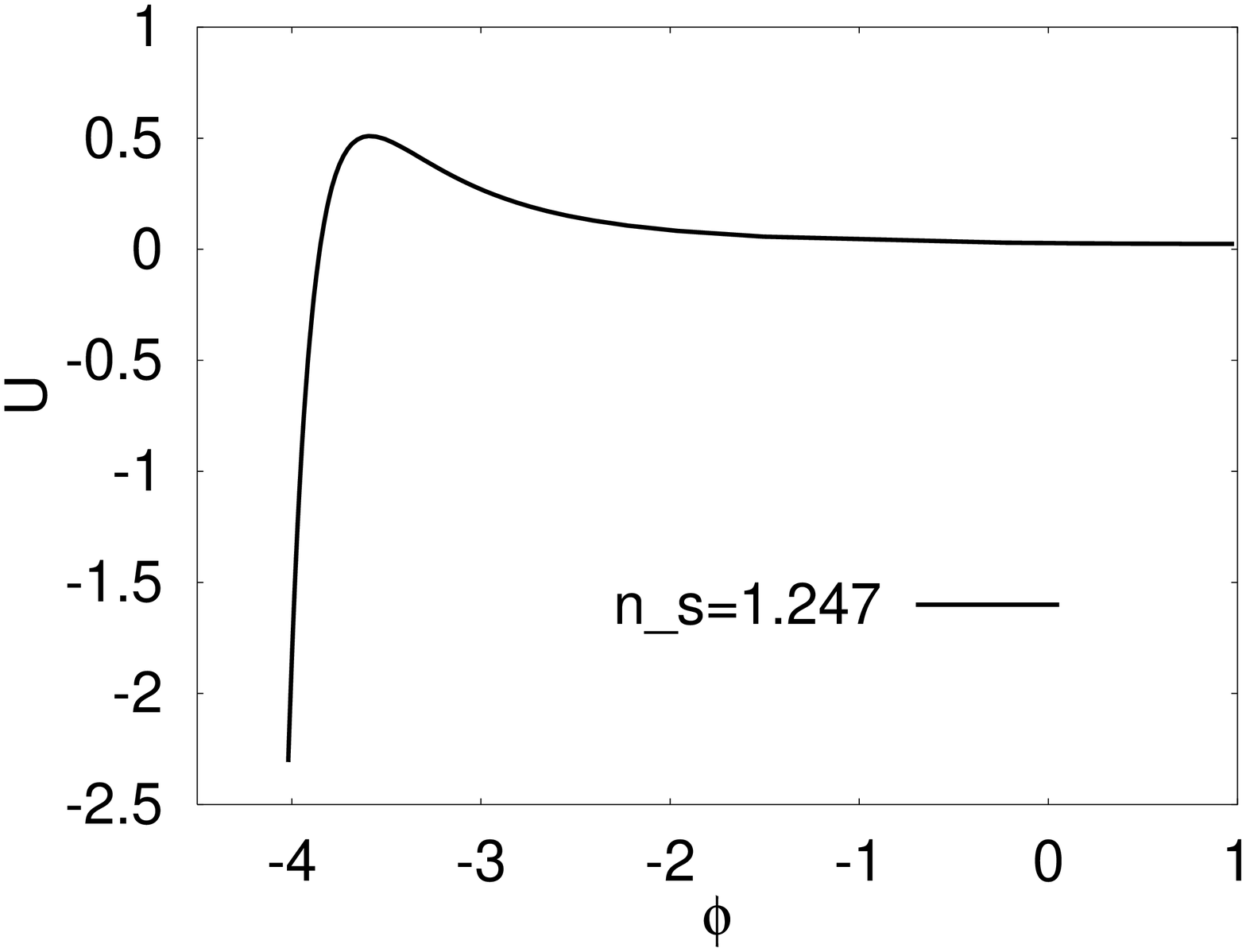} \hfill \\
\vskip.5cm
\caption[]{Inflationary potential $  U(\phi)$ with $n_s=1.247$ ($m=1$).
Further explanation in the Appendix.}
\label{pot1}
\end{figure}

In order to derive Fig.~\ref{pot1}, we used $A_2=0.01$ and found
$\bar y \approx 0.008026$ which is linked to $f(\bar y)\approx 2250$.
Furthermore, we applied the initial values $f'(\bar y)=-1$ and
$\phi(\bar y)=1$. The initial value for the potential is, of course,
$\kappa\widetilde U(\bar y)=3 \bar y$.
By this, the point $\phi=1$ in Fig.~\ref{pot1} is
correlated to $\kappa\widetilde U(H) = 3H^2$
(no potential does exist for $\phi>1$); inflation
($\kappa\widetilde U(H) > 2H^2$) occurs till almost the maximum of
the potential; for smaller values of $\phi$ kinetic energy
dominates and the scalar field becomes constant. 
We have investigated a broad interval of $f(\bar y)$
(between 0.1 and $10^6$) and $\phi(\bar y)$ values but always found this form
of the potential. For $A_2<0$, the function $f(y)$ is always negative,
hence, $\phi$ always imaginary. For $m=2$, we put $f(\bar y)=2250$ so that
$\bar y \approx 0.083427$; $f'(\bar y)=-1$ and $\phi(\bar y)=1$.

\newpage

\begin{figure}
\centering \leavevmode \epsfxsize=8cm \epsfbox{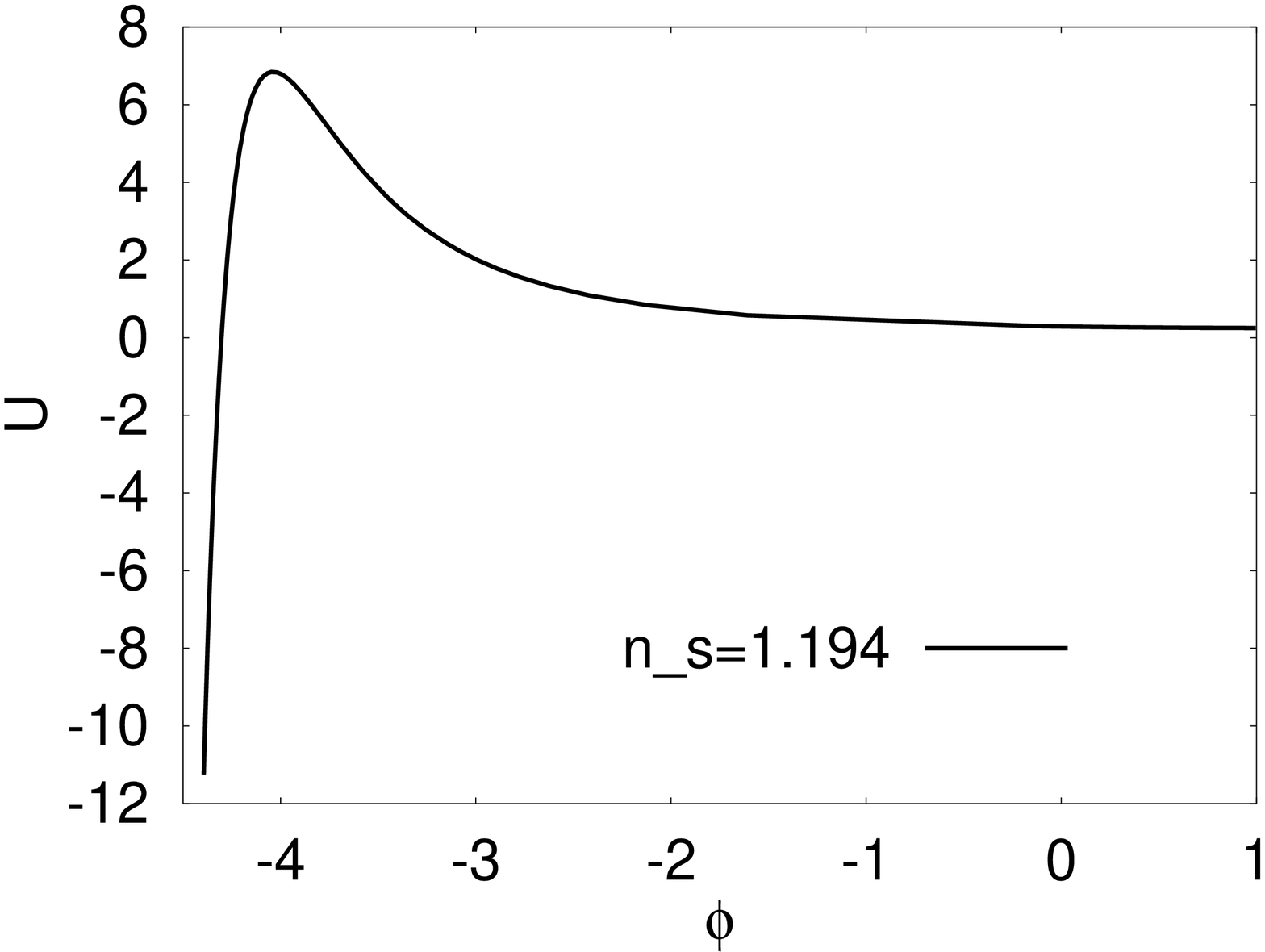} \hfill \\
\vskip.5cm
\caption[]{Inflationary potential $  U(\phi)$ with $n_s=1.194$ ($m=2$).
Further explanation in the Appendix.}
\label{pot2}
\end{figure}

\newpage

%--------------------------table beginning-------------------------  
\begin{table}
\caption{\em Eigenvalues of the spectral index}\label{table}
\begin{tabular}{lcc}
{\bf Eigenvalue} $n_s$&{\bf Spectrum }&{\bf Potential} $  U(\phi)$ \\ \hline
$(-6.57797)$ & (unrealistic) & $\sim\tan^2(\sqrt{\kappa}\phi/4)$ \\
$-3<n_s<1$ & continuous & $\propto \exp (\pm \sqrt{2 \kappa A_0}\; \phi )$ \\
\vdots       & discrete & \vdots \\
1.106   & ($m=5$) & \      \\
1.125   & All CMB data & \      \\
1.153   & ($m=3$) & \      \\
1.194   & COBE data & cf.~Fig.~\ref{pot2} \\
1.247   & ($m=1$) & cf.~Fig.~\ref{pot1}  \\
1.49575 & \ & $\sim \tan^2(\sqrt{\kappa}\phi/4)$ \\
\end{tabular}
\end{table}
%--------------------------table end-------------------------  

\end{document}